\documentclass[12pt]{article}

\usepackage{graphicx}
\usepackage[centertags]{amsmath}

\begin{document}
\newcommand{\be}{\begin{equation}}
\newcommand{\ee}{\end{equation}}
\newcommand{\bea}{\begin{eqnarray}}
\newcommand{\eea}{\end{eqnarray}}
\newcommand{\bib}{\bibitem}
\newcommand{\ita}{{\it et\, al}}
\author{S.S. Rajah $^{\footnote{Permanent address:
Department of Mathematics, Durban University of Technology,
Steve Biko Campus, Durban 4001, South Africa}}$ and S.D. Maharaj \\
Astrophysics and Cosmology Research Unit,\\
School of Mathematical Sciences, University of KwaZulu-Natal,\\ Private Bag X54001, Durban 4000, South Africa}
\title{A Riccati equation in radiative stellar collapse}
\maketitle
\begin{abstract}
We model the behaviour of a relativistic spherically symmetric
shearing fluid undergoing gravitational collapse with heat flux. It
is demonstrated that the governing equation for the gravitational
behaviour is a Riccati equation. We show that the Riccati equation
admits two classes of new solutions in closed form. We regain
particular models, obtained in previous investigations, as special
cases. A significant feature of our solutions is the general spatial
dependence in the metric functions which allows for a wider study of
the physical features of the model, such as the behaviour of the
causal temperature in inhomogeneous spacetimes.
\end{abstract}

\section{Introduction}
An important application of Einstein's general relativity theory in
relativistic astrophysics is the gravitational collapse of a
radiating star. Since the first idealized model of a static
spherical dust ball proposed by Oppenheimer and Snyder
\cite{oppenheimer}, many attempts have been made to describe more
realistic situations. The derivation of the junction conditions for
a radiating star with the exterior Vaidya metric was first obtained
by Santos \cite{santos}, and this formed  the basis on which simple
exact radiating models are constructed. The junction conditions were
later generalised by Chan {\ita}  \cite{chan1} to incorporate
pressure anisotropy, and by Maharaj and Govender \cite{maharaj1},
amongst others, to include the electromagnetic field. Particular
solutions of the Einstein equations and boundary conditions have
been found to describe the physical scenarios described above. These
solutions have been used to study the cosmic censorship hypothesis,
and various physical features including the adiabatic stability of
the model, surface luminosity, and the role of relaxational effects
on the thermal evolution in the causal thermodynamic theory.

In the absence of general techniques to analyse the nonlinear
coupled partial differential equations in Einstein's theory, many
formulations by researchers have sought to simplify the system of
equations by considering the special case of shearfree collapse. A
successful attempt at an exact model was made by Kolassis {\ita}
\cite{kolassis}, where the assumption of geodesic motion of the
fluid particles led to a considerable  simplification of the field
equations. Herrera {\ita} \cite{herrera1} solved Einstein's
equations, and reduced the junction condition at the boundary, for
shearfree collapse, to a nonlinear ordinary differential equation by
requiing that all Weyl tensor components vanish. It was initially
thought that this model could only be studied approximately. However
this conformally flat model was later solved exactly by Maharaj and
Govender \cite{maharaj2} by introducing a transformation that
linearised the boundary condition. Herrera {\ita} \cite{herrera2}
subsequently generalised the Maharaj and Govender \cite{maharaj2}
model and showed that other solutions to the linearised boundary
condition existed. In a recent treatment, Misthry {\ita}
\cite{misthry} extended the conformally flat model to an exact
nonlinear regime by transforming the governing equation to an Abel
equation of the first kind. These solutions may be used to study
realistic behaviour of the gravitating star with nonlinear boundary
conditions.

A natural extension of shearfree models is to include the effects of
shear and pressure anisotropy. The origins and effects of anisotropy
were first investigated by Herrera and Santos \cite{herrera3} and
later by Chan {\ita} \cite{chan2} and Herrera {\ita}
\cite{herrera4}. The formulation of a shearing model with pressure
anisotropy leads to a nonlinear partial differential equation at the
boundary; this presents a formidable difficulty in obtaining
solutions in closed form. Earlier treatments have been qualitative
in nature without exact analytical solutions, and hence researchers
have used numerical techniques to study the physical behaviour of
the model. Noguiera and Chan \cite{noguiera}, assuming separable
forms for the metric functions, reduced the boundary condition to a
nonlinear ordinary differential equation. They then investigated the
equation numerically and a detailed study of the physical features
of the model was performed. Recently Naidu {\ita} \cite{naidu}
obtained the first exact analytical model with nonzero shear by
considering geodesic motion of the fluid particles. Their particular
solution, however, has singularities at the stellar core. Maharaj
and Misthry \cite{maharaj3} obtained two classes of exact solutions,
nonsingular at the centre, in which the Naidu {\ita} \cite{naidu}
model is contained as a special case. Note that the  spatial
components of the metric functions have been restricted to
particular forms in the Maharaj and Misthry \cite{maharaj3} models.
This restriction limits the investigation of the relativistic
effects such as the relaxation time scales on the model, and
consequently  it is desirable to obtain a wider class of solutions.

The main objective of this paper is to carry out a systematic study
of the governing equation, at the boundary, for the shearing
collapse of a compact radiating stellar fluid model. We seek to
obtain a general class of nonsingular exact solutions which allows
for flexibility in the choice of physical parameters required to
investigate the physical features of the model. It is not desirable
to eliminate the inherent nonlinearity at the boundary; instead we
seek to transform the governing equation to a familiar form, namely
the Riccati equation. In Section 2, we formulate the model using the
Einstein field equations together with the junction conditions. In
Section 3, we first transform the governing equation into a Riccati
equation, and then propose two transformations which lead to
separable equations. Two new classes of exact solutions are found.
It is shown that solutions found earlier with nonzero shear are
contained in our models. In Section 4, we integrate the truncated
form of the Maxwell-Cattaneo heat transport equation and obtain an
explicit form for the causal temperature. We present profiles for
the causal and acausal temperatures and briefly discuss the features
of the graphs.

\section{Formulation of the model}
We seek to model a spherically symmetric star undergoing radiative
gravitational collapse with nonzero shear in the context of general
relativity. The line element describing the gravitational field for
the interior spacetime is taken to be \be
ds^2=-dt^2+B^2dr^2+Y^2(d\theta^2 +\sin^2\theta d\phi^2)\label{intl}
\ee where $B$ and $Y$ are functions of both the temporal coordinate
$t$ and radial coordinate $r$. For this model, the fluid
four-velocity $u^a=\delta_0^a$ is comoving. The fluid
four-acceleration vector $\dot u^a$, the expansion scalar $\Theta$,
and the magnitude of the shear scalar $\sigma$, respectively, are
given by
\begin{subequations}
\label{fluidp}
\bea
\dot u^a&=&0\\
\Theta &=& \left(\frac{\dot{B}}{B}+2\frac{\dot Y}{Y}\right)\\
\sigma &=&\frac{1}{3}\left(\frac{\dot Y}{Y}-\frac{\dot B}{B}\right)
\eea
\end{subequations}
for the line element (\ref{intl}). We observe that the particle
trajectories of the collapsing fluid are geodesics because $\dot
u^a=0$. However note from (\ref{fluidp}) that the fluid expansion
and the shear may be nonzero in general. A similar analysis was
performed by Kolassis $\ita$  \cite{kolassis} when the shear is
vanishing; in that case it was possible to solve the boundary
condition and the field equations and obtain an exact solution. The
Kolassis $\ita$  \cite{kolassis} model has a Friedmann limit and the
dust cosmological model is regained. It would be interesting to
compare the temporal evolution of the model when shear is present.
Investigation of the behaviour of the temperature in causal
thermodynamics for geodesic motion has revealed  higher central
temperatures than the Eckart theory as established by Govender
$\ita$ \cite{govender1}. In this study we seek to incorporate the
effects of shear in the model.

The energy momentum tensor for the interior spacetime has the form
\be
T_{ab}=(\rho+p)u_au_b+pg_{ab}+q_au_b+q_bu_a+\pi_{ab}\label{momt}
\ee where $\rho$ is the density of the gravitating fluid, $p$ is
the isotropic pressure, $q_a$ is the heat flux vector and
$\pi_{ab}$ is the stress tensor. These quantities are measured
relative to the four-velocity $\textit{\textbf{u}}$. The stress
tensor can be written explicitly as
\be \pi_{ab}=(p_r-p_t)\left(n_a
n_b-\frac{1}{3}h_{ab}\right) \ee where $p_r$ is the radial
pressure, $p_t$ is the tangential pressure, and
$\textit{\textbf{n}}$ is the unit radial vector orthogonal to
$\textit{\textbf{u}}$. Hence we have
$n^a=\frac{1}{B}\delta_1^a.$ The isotropic pressure \be
p=\frac{1}{3}(p_r+2p_t) \ee relates the radial pressure and the
tangential pressure.

It is possible to write the Einstein field equations as the set
\begin{subequations}\label{EFE}
\bea
\rho&=&2\frac{\dot B}{B}\frac{\dot Y}{Y}+\frac{1}{Y^2}
+\frac{\dot Y^2}{Y^2}-\frac{1}{B^2}\left(2\frac{Y''}{Y}
+\frac {Y'^2}{Y^2}-2\frac{B'}{B}\frac{Y'}{Y}\right)\\
\nonumber\\
p_r&=&\left(-2\frac{\ddot Y}{Y}-\frac{\dot Y^2}{Y^2}+2\frac{\dot Y}{Y}\right)
+ \frac{1}{B^2}\left(\frac{Y'^2}{Y^2}\right)-\frac{1}{Y^2}\\
\nonumber\\
p_t&=&\frac{1}{B^2}\left(-\frac{B'}{B}\frac{Y'}{Y}+\frac{Y''}{Y}\right)
-\left(\frac{\ddot B}{B}+\frac{\dot B}{B}\frac{\dot Y}{Y}+\frac{\ddot Y}{Y}\right)\\
\nonumber\\
q&=&-\frac{2}{B^2}\left(-\frac{\dot Y'}{Y}+\frac{\dot B}{B}\frac{Y'}{Y}\right)
\eea
\end{subequations}
for the spacetime (\ref{intl}) and the matter distribution
(\ref{momt}).  The fluid pressure is anisotropic and the
heat flux $q^a=(0,q,0,0)$ has only a
nonvanishing radial component. We observe that if functional forms for the
gravitational potentials $B$ and $Y$ are given then expressions for the matter
variables $\rho, p_r, p_t$ and $q$ immediately follow from
(\ref{EFE}).

The exterior spacetime, describing the region outside the
stellar boundary, is described by the Vaidya metric

\be
ds^2=-\left(1-\frac{2m(v)}{\text{R}}\right)dv^2-2dv d\text R +
\text R^2(d\theta^2+\sin^2\theta d\phi^2) \label{vaidya}
\ee
\\
where $m(v)$ denotes the mass of the fluid as measured by an
observer at infinity. The metric (\ref{vaidya}) describes
coherent null radiation which is flowing in the radial direction relative to the
hypersurface $\Sigma$ which represents the boundary of the star.
The matching of the exterior spacetime with the interior spacetime leads to the
following set of junction conditions on the hypersurface $\Sigma$:
\begin{subequations}\label{junct}
\bea
dt&=&\left(1-\frac{2m}{\text R_\Sigma}+2\frac{d\text R_\Sigma}{dv}
\right)^{1/2}dv \label{juncta}\\
Y(\text R_\Sigma,t)&=&\text R_{\Sigma}(v) \label{junctb}\\
m(v)_\Sigma &=&\left[\frac{Y}{2}\left(1+\dot Y^2-\frac{Y'^2}{B^2}
\right)\right]_\Sigma\label{junctd}\\
(p_r)_{\Sigma}&=&(qB)_{\Sigma}\label{junctc}
\eea
\end{subequations}
The nonvanishing of the radial pressure $p_r$ at $\Sigma$ leads to
an additional differential equation, namely the boundary condition
(\ref{junctc}), which has to be satisfied together with the field
equations (\ref{EFE}). This condition was first established by
Santos \cite{santos} for shearfree spacetimes, and extended to
spacetimes with nonzero shear by Glass \cite{glass} and Maharaj and
Govender \cite{maharaj4}, amongst others. In a recent investigation
by Di Prisco $\ita$ \cite{prisco1}, the matching conditions
applicable to spherically symmetric gravitational collapse with
dissipation and nonzero shear have been generalised to include
nonadiabatic charged fluids.

\section{Solution of the governing equation}

The junction condition $(p_r)_\Sigma =(qB)_{\Sigma}$ becomes
 \be
2Y\ddot{Y}+\dot{Y}^2-\frac{Y'^2}{B^2}+\frac{2}{B}Y\dot{Y}'
-2\frac{\dot{B}}{B^2}YY'+1=0 \label{junceqn} \ee which follows from
(\ref{EFE}). Equation (\ref{junceqn}) governs the gravitational
behaviour of a radiating star with anisotropic pressure and nonzero
shear. To complete the description of the gravitational behaviour of
the model we need to integrate the junction condition
(\ref{junceqn}); this will lead to functional forms for the metric
functions $B(r,t)$ and $Y(r,t)$. Exact solutions for the junction
condition (\ref{junceqn}) have been extremely difficult to obtain
due to the nonlinear nature of the equation. In an earlier study of
a shearing radiating model, Noguiera and Chan \cite{noguiera} used
numerical techniques to obtain approximate solutions. Ideally an
exact solution is desirable in terms of elementary or special
functions. An exact solution for this physical model was obtained by
Naidu {\ita} \cite{naidu} in terms of the elementary functions. This
class of solution is singular at the stellar centre. Maharaj and
Misthry \cite{maharaj3} extended the Naidu $\ita$ \cite{naidu} model
and showed that a wider category of solutions are possible; the
singularities at the centre were shown to be avoidable.

We seek, in this treatment, to obtain a general class of nonsingular
solutions which will allow for an investigation of the physical
features of the model. Previous treatments were ad hoc. Our
objective is to write (\ref{junceqn}) in a generic form, and then
obtain solutions systematically. It is shown that the model leads to
the formation of a Riccati equation where the potential $B$ is the
dependent variable. We present, in the following, a method of
solving (\ref{junceqn}) exactly, and find several classes of
solutions depending on the form of $Y$ used. We rewrite
(\ref{junceqn}) in the form \be
\dot{B}=\left(\frac{\ddot{Y}}{Y'}+\frac{\dot{Y}^2}{2YY'}
+\frac{1}{2YY'}\right)B^2+\frac{\dot{Y'}}{Y'}B-\frac{Y'}{2Y}
\label{Reqn1} \ee which is a Riccati equation in the potential $B$.
We demonstrate that two classes of solutions can be found for this
Riccati equation.

\subsection{The first solution}
We seek solutions where $Y$ is a separable function of the form \be
Y=R(r)(t+a)^{2/3} \label{Y1} \ee so that the temporal evolution of
the model is specified. It is convenient at this point to introduce
the transformation \be B=Z(t+a)^{2/3} \ee Then (\ref{Reqn1}) can be
written in the form \be (t+a)^{2/3}\dot Z
=\frac{1}{2RR'}(Z^2-R'^2)\label{transeqn1} \ee Equation
(\ref{transeqn1}) is simple and separable with solution \be
Z=R'\left(\frac{1+g(r)\exp[3(t+a)^{1/3}/R]}{1-g(r)\exp[3(t+a)^{1/3}/R]}\right)
\ee where $g(r)$ is related to an arbitrary constant of integration.
Consequently the potential $B$ can be obtained explicitly in the
form \be B=R'\left(\frac{1+g(r)\exp[3(t+a)^{1/3}/R]}{1-g(r)\exp[3(t
+a)^{1/3}/R]}\right)(t+a)^{2/3}\label{B1} \ee
 From (\ref{Y1}) and
(\ref{B1}) we may write the interior metric (\ref{intl}) as \be
ds^2=-dt^2+(t+a)^{4/3}\left[R'^2\left(\frac{1+g(r)
\exp[3(t+a)^{1/3}/R]}{1-g(r)\exp[3(t+a)^{1/3}/R]}\right)^2dr^2+R^2(d\theta^2+\sin
^2\,\theta d\phi^2)\right]\label{intsol1} \ee which describes the
interior spacetime of the radiating star.

The matter variables for the model are given by

\begin{subequations}\label{EFE1}
\bea
\rho&=&\frac{4}{3\tilde{t}\,^2}\left[1+\frac{3
\tilde t\,^{2/3}}{4R^2}\right]
-\left[\frac{1-g\exp[{3\tilde t\,^{1/3}/R}]}{R
\tilde t\,^{2/3}(1+g\exp[{3\tilde t\,^{1/3}/R}])}\right]^2\nonumber\\
&&+\frac{8g\exp[{3\tilde t\,^{1/3}/R}]}{3 \tilde
t\,^{5/3}R(1-g^2\exp{2[{3\tilde t\,^{1/3}/R}]})}\nonumber\\
&&+\frac{4\exp[{3\tilde t\,^{1/3}/R}](1-g \exp[{3\tilde
t\,^{1/3}/R}])}{RR'\tilde t(1+g[ \exp{3\tilde
t\,^{1/3}/R}])^3}\left[\frac{g'}{
\tilde t\,^{1/3}}-\frac{3g}{R^2}\right]\\
 p_r&=&-\frac{4g\exp[{3\tilde t\,^{1/3}/R}]}{
\tilde t\,^{4/3}R^2(1+g\exp[{3\tilde t\,^{1/3}/R}]^2}\\
 p_t&=&-\frac{2g\exp[{3\tilde t\,^{1/3}/R}]}{ R\,^2\tilde
t\,^{4/3}(1-g^2\exp2[{3\tilde t\,^{1/3}/R}])}\left[1
+\frac{2g\exp[{3\tilde t\,^{1/3}/R}]}{(1-g\exp[{3
\tilde t\,^{1/3}/R}])}+\frac{4R}{3\tilde t\,^{1/3}}\right]\nonumber\\
 &&-\frac{2\exp[{3\tilde t\,^{1/3}/R}](1-g\exp{[{ 3\tilde
t\,^{1/3}/R}]})}{R R\,'\tilde t(1+g\exp{[{ 3\tilde
t\,^{1/3}/R}]})^3}\left[\frac{g'}{
\tilde t\,^{1/3}}-\frac{3g}{R\,^2}\right]\\
 q&=&-\frac{4g\exp[{3\tilde t\,^{1/3}/R}](1-g\exp{[{ 3\tilde
t\,^{1/3}/R}]})}{R\,^2 R'\tilde t\,^2(1+ g\exp{[{3\tilde
t\,^{1/3}/R}]})^3} \eea
\end{subequations}
which satisfies the Einstein equations (\ref{EFE}).
For simplicity we have set $\tilde t = t+a.$

We have obtained an exact solution to the Einstein field
equations (\ref{EFE}) that satisfies the boundary condition
(\ref{junceqn}) for a radiating relativistic star.
This is a new class of exact solutions where the
spatial dependence in the function $R$ is arbitrary.
Consequently particular solutions found in the past
can be shown to be contained in this class. We observe
that when $R=r+b$ the metric (\ref{intsol1}) becomes
\bea
ds^2&=&-dt^2+(t+a)^{4/3}\left[\left(\frac{1+g(r)\exp[3
(t+a)^{1/3}/(r+b)]}{1-g(r)\exp[3(t+a)^{1/3}/
(r+b)]}\right)^2dr^2 \right. \nonumber \\
&&+\left.(r+b)^2(d\theta^2+\sin ^2\,\theta
d\phi^2)\right]\label{metric1} \eea
 Thus we have regained the
Maharaj and Misthry \cite{maharaj3} model which is regular at the
stellar origin. The Naidu $\ita$ \cite{naidu} model is regained from
(\ref{intsol1}) when $a = 0$ and $b = 0$. Other choices for the
function $R$ are clearly possible: the choice should be such that
the model remains regular at the centre and the model is well
behaved in the interior. We further observe that the model yields
the Friedmann dust model when $g = 0$. In this case we can find
coordinates such that \be ds^2=-dt^2+t^{4/3}[dr^2+r^2(d\theta^2
+\sin^2\theta d\phi^2)] \ee for which the heat flux vector $q_a$
vanishes and $p_r = p_t =0$ with $\rho=\frac{4}{3t^2}$ in geodesic
motion.

\subsection{The second solution}
Other solutions to the Riccati equation (\ref{Reqn1}) exist in
closed form but these are difficult to find in practice. It is
possible to find a second class of solutions to (\ref{Reqn1}) by
assuming that \be Y=R(r)(t+a)\label{Y2} \ee
 In this case we
introduce the transformation \be B=(t+a)Z \ee Then (\ref{Y2}) can be
written in the form \be (t+a)\dot Z=\frac{R^2+1}{2RR'}\left(Z^2-
\frac{R'^2}{R^2+1}\right)\label{transeqn2} \ee Equation
(\ref{transeqn2}) is a simple equation, separable in the variables
$Z$ and $t$, with solution \be
Z=\frac{R'}{\sqrt{R^2+1}}\left(\frac{1+h(r)(t
+a)^{^{\sqrt{\frac{R^2+1}{R^2}}}}}{1-
h(r)(t+a)^{^{\sqrt{\frac{R^2+1}{R^2}}}}}\right) \ee where $h(r)$ is
an arbitrary function of integration. Therefore we can express the
metric potential in the form \be
B=\frac{R'}{\sqrt{R^2+1}}\left(\frac{1+h(r)(t
+a)^{^{\sqrt{\frac{R^2+1}{R^2}}}}}{1-h(r)(t
+a)^{^{\sqrt{\frac{R^2+1}{R^2}}}}}\right)(t+a)\label{B2} \ee Then
from (\ref{Y2}) and (\ref{B2}) we obtain the interior metric \be
ds^2=-dt^2+(t+a)^2\left[\frac{R'^2}{R^2+1}\left(
\frac{1+h(r)(t+a)^{^{\sqrt{\frac{R^2+1}{R^2}}}}}{
1-h(r)(t+a)^{^{\sqrt{\frac{R^2+1}{R^2}}}}}\right)^2dr^2
+R^2(d\theta^2+\sin ^2\,\theta d\phi^2)\right] \ee which describes
the stellar interior.

The Einstein field equations (\ref{EFE}) then imply the
 matter variables which are given by
\begin{subequations}
\bea
\rho&=&\frac{2}{\tilde t\,^2}\left[1+\frac{1}{2R^2}
\right]-(3R^2+1)\left[\frac{1-h\tilde t\,^{^{\sqrt{
\frac{R^2+1}{R^2}}}}}{R\tilde t\left(1+h\tilde t\,^{^{
\sqrt{\frac{R^2+1}{R^2}}}}\right)}\right]^2\nonumber\\
&&+\frac{2h\sqrt{R^2+1}}{RR'}\frac{\tilde t\,^{^{\sqrt{
\frac{R^2+1}{R^2}}}}}{\tilde t\,^2\left(1-h^2\,\tilde t\,^{^{2\sqrt{
\frac{R^2+1}{R^2}}}}\right)}\nonumber\\
&&+\frac{4\left(R^2+1\right)^2\,\tilde t\,^{^{\sqrt{
\frac{R^2+1}{R^2}}}}}{R'R\tilde t\,^2\left(1+h\,
\tilde t\,^{^{\sqrt{\frac{R^2+1}{R^2}}}}\right)^2}
\left[\frac{h'}{R^2+1}-\frac{h\ln{\tilde t}}{R^9}\right]\\
 p_r&=&\frac{1}{\tilde t\,^2}\frac{R^2+1}{R^2}\left[
\left(\frac{1-h\tilde t\,^{^{\sqrt{\frac{R^2+1}{R^2}}}}}{
1+h\tilde t\,^{^{\sqrt{\frac{R^2+1}{R^2}}}}}\right)^2-1\right]\\
 p_t&=&-\frac{2h\sqrt{R^2+1}\tilde t\,\,^2{^{\sqrt{
\frac{R^2+1}{R^2}}}}}{R\tilde t\,^2\left(1-h^2\, \tilde
t\,\,^2{^{\sqrt{\frac{R^2+1}{R^2}}}}\right)}\left[1
+\frac{\sqrt{R^2+1}}{RR'}-\frac{1}{R'}\right]-\frac{1}{
\tilde t\,^2}\nonumber\\
&&+\frac{1}{\tilde t\,^2}\left(\frac{1-h\,\tilde t\,^{^{
\sqrt{\frac{R^2+1}{R^2}}}}}{1+h\,\tilde t\,^{^{\sqrt{
\frac{R^2+1}{R^2}}}}}\right)^2-\frac{4h^2(R^2+1)\,
\tilde t\,^{^{\sqrt{\frac{R^2+1}{R^2}}}}}{R^2R'
\tilde t\,^2\left(1-h^2\,\tilde t\,\,^2{^{\sqrt{
\frac{R^2+1}{R^2}}}}\right)\left(1-h\,\tilde t\,^{^{\sqrt{\frac{R^2+1}{R^2}}}}\right)}
\nonumber\\
&&-\frac{2(R^2+1)\tilde t\,^{^{\sqrt{\frac{R^2+1}{R^2}}}}
\left(1-h\tilde t\,^{^{\sqrt{\frac{R^2+1}{R^2}}}}\right)}{RR'
\tilde t\,^2\left(1+h\tilde t\,^{^{\sqrt{\frac{R^2+1}{R^2}}}}
\right)^2}\left[h'-\frac{h(R^2+1)\ln \tilde t}{R^9\left(1+h\,
\tilde t\,^{^{\sqrt{\frac{R^2+1}{R^2}}}}\right)}\right]\\
 q&=&-\frac{4h(R^2+1)^{3/2}\tilde t\,^{^{
\sqrt{\frac{R^2+1}{R^2}}}}\left(1-h\tilde t\,^{^{
\sqrt{\frac{R^2+1}{R^2}}}}\right)}{R^2R'^2\left( 1+h\tilde
t\,^{^{\sqrt{\frac{R^2+1}{R^2}}}}\right)^3} \eea
\end{subequations}
where we have set $\tilde t = t+a$ for convenience.

We have generated another exact solution to the field equations
(\ref{EFE}) that satisfies the boundary condition (\ref{junceqn}).
Again the spatial dependence in the function is
arbitrary and the model is regular at the centre.
We observe that the solution contains an exponential
temporal dependence on the spatial function $R(r)$ of
the form $t^{\sqrt{(R^2+1)/R^2}}$. Such solutions are
difficult to interpret but may be relevant in describing
new physical models in the strong gravity regime for gravitational collapse.

\section{Causal temperature}
The simple forms of the solutions found in this paper, in particular
the first solution, permit a study of the physical features. We
consider briefly the relativistic effects on the temperature. For a
shearing superdense matter distribution, we employ the
Maxwell-Cattaneo heat transport equation to investigate the causal
thermodynamical behaviour of the model. In the absence of rotation
and viscous stress this is given by the truncated version \be \tau
h_a^b\dot q_b+q_a=-\kappa(h_a^b\nabla_bT+T\dot u_a)\label{M-C} \ee
where $\tau$ is the relaxation time, $\kappa$ is the thermal
conductivity, and  $h_{ab}=g_{ab}+u_au_b$ projects into the comoving
rest space. When $\tau=0$ we regain the acausal Fourier heat
transport equation. For our model equation (\ref{M-C}) may be
written as \be T=-\frac{\tau}{\kappa}\int (qB)^{^\textbf .}Bdr-
\frac{1}{\kappa}\int qB^2dr \label{Temp} \ee describing the
evolution of the causal temperature. It has been shown in previous
investigations of relativistic stellar models that the effect of the
relaxation time $\tau$, on the thermal evolution, plays a
significant role in the latter stages of collapse \cite{martinez,
govender2, govender3, prisco2}. Naidu {\ita} \cite{naidu} showed
that in the presence of shear stress, the relaxation time decreases
as the collapse proceeds and the central temperature increases. The
particular form of the relaxational time $\tau$ is dependent on the
physical constraints of the model during the latter phases of
collapse. We observe that since our solutions have elementary
functions with an arbitrary form for the spatial component, it is
possible to integrate (\ref{Temp}) for different choices of $\tau$.
In particular, the effect of decreasing relaxation time with
decreasing radius and higher central temperature is possible by
incorporating a varying function $\tau$.

In this study we set $\tau$ and $\kappa$ to be constant in
order to examine the causal and acausal behaviour in the first
solution (\ref{metric1}). We need to choose particular forms
for the arbitrary function $R(r)$ to complete the integration.
As a first example we set $R=r+b$ to obtain for the temperature
\bea
T&=&\frac{4\tau}{9\kappa \tilde t\,^2}\left(\text{Log}
\left[\frac{(1+g\exp[3\tilde t\,^{1/3}/r+b])^3}{(-1+g
\exp[3\tilde t\,^{1/3}/r+b])^2}\right]\right)\nonumber\\
&&-\frac{4\tau}{3\kappa \tilde t\,^{5/3}}\left(\frac{g
\exp[3\tilde t\,^{1/3}/r+b]}{(r+b)(1+g\exp[3\tilde
t\,^{1/3}/r+b])}\right)-\frac{4}{3\kappa \tilde t}
\text{Tanh}^{-1}(g\exp[3\tilde t\,^{1/3}/r+b])+f(t)\label{temp1}
\eea
where we have kept $g$ constant, and $f(t)$ is a constant
of integration related to luminosity as observed by a distant
observer. As a second example we set $R=e^r$ to obtain the temperature
\bea
T&=&\frac{4\tau}{9\kappa \tilde t^2}\left(\text{Log}\left[
\frac{(1+g\exp[3\tilde t\,^{1/3}/e^r])^3}{(-1+g\exp[3\tilde
t\,^{1/3}/e^r])^2}\right]\right)\nonumber\\
&&-\frac{4\tau}{3\kappa \tilde t\,^{5/3}}\left(\frac{g\exp[3 \tilde
t\,^{1/3}/e^r]}{e^r(1+g\exp[3\tilde t\,^{1/3}/e^r])}
\right)-\frac{4}{3\kappa \tilde t}\text{Tanh}^{-1}(g\exp[3 \tilde
t\,^{1/3}/e^r])+f(t)\label{temp2} \eea
 Consequently it is possible
to find analytic forms for the causal temperature in terms of
elementary functions as shown in (\ref{temp1}) and (\ref{temp2}). We
regain the noncausal (Eckart) temperature profiles when $\tau =0$.
Our simple forms for $T$ assist in studying the evolution of a
radiating star in different time intervals. These models provide
examples of temperatures where inhomogeneity is directly related to
dissipation.

\begin{figure}[h]
    \centering
    \includegraphics{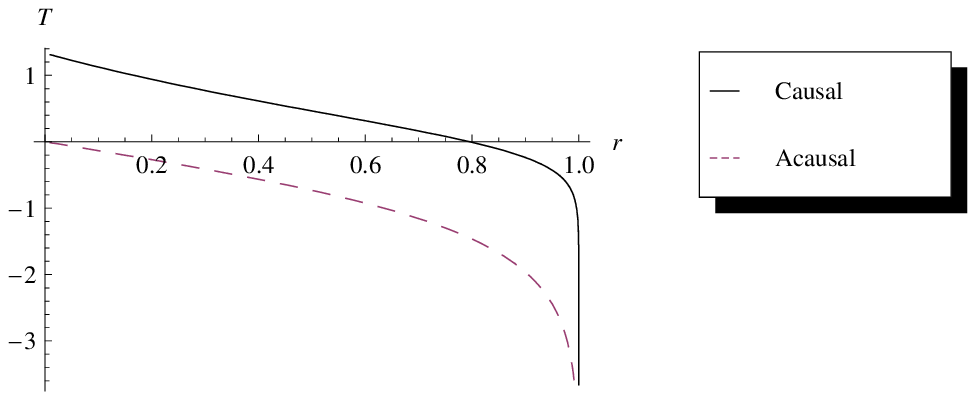}
    \caption{Temperature $T$ vs radial coordinate $r$ $(\tau =1)$}
 \label{fig:temp1}
\end{figure}

\begin{figure}[h]
    \centering
    \includegraphics{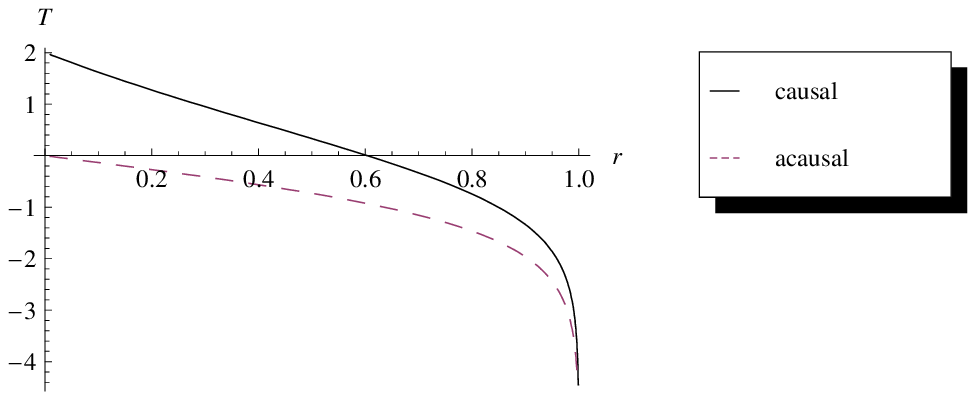}
    \caption{Temperature $T$ vs radial coordinate $r$ $(\tau > 1)$}
    \label{fig:temp2}
\end{figure}

It is possible to qualitatively distinguish the causal and acausal
temperatures for the region between the centre and the surface of
the star. In Figure \ref{fig:temp1}, we provide plots of the causal
(solid line) and Eckart (broken line) temperatures against the
radial coordinate on the interval $0\le r \le 1$, where we have
selected $\tau =1$ for simplicity. This figure has been generated
with the help of Mathematica. We observe that the temperature is a
monotonically decreasing function as we approach the boundary from
the stellar centre. Also, it is immediately clear that the causal
temperature is everywhere greater than the acausal temperature in
the interior of the star. At the boundary $\Sigma$, however \be
T(t,r_\Sigma)_{causal}=T(t,r_\Sigma)_{acausal} \ee
 This simple figure
has been generated by assuming a particular constant value for the
relaxation time $\tau$ and the thermal conductivity $\kappa$.
 Changing the magnitude of these values would produce a change in
the separation of the curves but the results do not change
qualitatively. For example, in Figure \ref{fig:temp2}, we provide a
plot of the causal (solid line) and Eckart (broken line)
temperatures for $\tau
>0$. We note that in this case both temperatures decrease more
rapidly as we approach the boundary; the value of $\tau$ affects the
gradient of the temperature. As indicated previously it is possible
for the relaxation time $\tau$ to vary. The choice for $\tau$ should
be dictated on physical grounds, e.g. rate of particle production at
the stellar surface.

\vspace{0.5cm}

\vspace{0.5cm}

\section*{Acknowledgements}
SSR thanks the National Research Foundation and the Durban
University of Technology for financial support. SDM acknowledges
that this work is based upon research supported by the South
African Research Chair Initiative of the Department of Science and
Technology and National Research Foundation.

\end{document}